\documentclass[11pt]{article}

\usepackage[english]{babel}

\usepackage{graphicx}
\usepackage[colorlinks=true, allcolors=blue]{hyperref}

\usepackage{amssymb,amsmath,amsthm,amsfonts,amstext}
\usepackage[left=1.5cm,right=1.5cm,top=1.5cm,bottom=1.5cm]{geometry}
\usepackage{enumerate}

\usepackage[mathscr]{euscript}

\usepackage[title]{appendix}

\usepackage{thm-restate}	%re-stating results in Appendix

\usepackage{blkarray}		%displaying matrix A in sec.4

\newtheorem{theorem}{Theorem}

\newtheorem{proposition}[theorem]{Proposition}
\newtheorem{lemma}[theorem]{Lemma}

\newtheorem{claim}[theorem]{Claim}

\newcommand\IGNORE[1]{}

\newcommand{\redtext}[1]{{\color{red}{#1}}}
\newcommand{\bluetext}[1]{{\color{blue}{#1}}}

\newcommand{\browntext}[1]{{\color{brown}{#1}}}
\newcommand{\tealtext}[1]{{\color{teal}{#1}}}

\usepackage{tikz}
\usetikzlibrary{decorations}
\usetikzlibrary{decorations.pathreplacing}

\newcommand*\circled[1]{\tikz[baseline=(char.base)]{
    \node[shape=circle,draw,inner sep=2pt] (char) {#1};}}

\newcommand\lpopt{\hbox{\text{LP}$_{\text{opt}}$}}

\newcommand{\Qp}{\ensuremath{\Q_{\geq 0}}}
\newcommand{\Zint}{\ensuremath{\mathbb Z}}
\newcommand{\Zp}{\ensuremath{\Zint_{\geq 0}}}

\newcommand{\C}{\mathscr{C}}	%% script C
\newcommand{\J}{{J}}		%% edge-set J

\newcommand\capbound{\lambda}

\newcommand{\pliableapx}{6}

\newcommand\ASC{\mathrm{Cover\,Small\,Cuts}}

\newcommand{\ds}{\displaystyle}
\newcommand\N{N}
\newcommand\Q{Q}
\newcommand\q{q}
\newcommand{\p}{\phi}
\newcommand{\hp}{{\phi}}
\newcommand{\bL}{\overline{L}}
\newcommand{\hL}{\widehat{L}}
\title{A Bad Example for Jain's Iterative Rounding Theorem for the Cover~Small~Cuts Problem}

\author{\large
Miles Simmons\thanks{
        {\tt mjsimmons@uwaterloo.ca}.
        Department of Combinatorics \& Optimization, University of Waterloo, Canada.}
\and 
Ishan Bansal\thanks{
        {\tt ib332@cornell.edu}.
	Amazon, Bellevue, WA, USA. This work is external and does not relate to the position at Amazon. }
\and
Joseph Cheriyan\thanks{
{\tt jcheriyan@uwaterloo.ca}.
        Department of Combinatorics \& Optimization, University of Waterloo, Canada.}
}

\begin{document}
\maketitle

%%%%%%%%%%%%%%%%%%%%%%%%%%%%%%%%%%%%%%%%%%%%%%%%%%%%%%
\begin{abstract}
{
Jain's iterative rounding theorem is a well-known result in
the area of approximation algorithms and, more broadly, in combinatorial optimization.
The theorem asserts that LP relaxations of several problems in
network design and combinatorial optimization
have the following key property:
for every basic solution $x$ there exists a variable $x_e$ that has
value at least a constant (e.g., $x_e\geq\frac12$).

We construct an example showing that this property fails to hold for the $\ASC$ problem.
In this problem, we are given an undirected, capacitated graph $G=(V,E),u$
and a threshold value $\lambda$,
as well as a set of links $L$ with end-nodes in $V$
and a non-negative cost for each link $\ell\in L$;
the goal is to find a minimum-cost set of links such that
each non-trivial cut of capacity less than $\lambda$ is covered by a link.

This indicates that the polyhedron of feasible solutions to the LP (for $\ASC$)
differs in an essential way from the polyhedrons associated with several problems
in combinatorial optimization.
Moreover, our example shows that a direct application of Jain's iterative rounding algorithm
does not give an $O(1)$ approximation algorithm for $\ASC$.
We mention that Bansal~et~al.~\cite{BCGI24} present an $O(1)$~approximation algorithm
for $\ASC$ based on the primal-dual method of Williamson~et~al.~\cite{WGMV95}.
}
\end{abstract}
%%%%%%%%%%
{
\section{Introduction \label{sec:intro}}

Jain's iterative rounding theorem \cite{Jain01} is a well-known result in the area of
approximation algorithms and, more broadly, in combinatorial optimization \cite{LRS-book}.
The theorem asserts that LP relaxations of several problems in
network design and combinatorial optimization
have the following key property:
for every basic solution $x$ there exists a variable $x_e$ that has
value at least a constant (e.g., $x_e\geq\frac12$).

We construct an example showing that this property fails to hold for the $\ASC$ problem.

%%%%%
%	\subsubsection{The \texorpdfstring{$\ASC$}{Cover Small Cuts} problem}
\subsection{The {$\ASC$} problem}
{
We follow the notation from \cite[Section~1.3]{BCGI:arxiv}.
In an instance of the $\ASC$ problem, we are given an
undirected capacitated graph $G = (V,E)$ with edge-capacities $u
\in \Qp^E$, a set of links $L \subseteq \binom{V}{2}$ with costs
$c \in \Qp^L$, and a threshold $\capbound \in \Qp$.
A subset $F \subseteq L$ of links is said to \emph{cover} a
node-set $S$ if there exists a link $e \in F$ with exactly one
end-node in $S$.
The objective is to find a minimum-cost $F\subseteq{L}$ that covers
each non-empty $S \subsetneq V$ with $u(\delta_E(S)) < \capbound$.

Let $\C = \{ \emptyset \neq S \subsetneq V : u(\delta_E(S)) < \capbound \}$.
Then we have the following covering~LP relaxation of the problem.

\begin{align*}\label{LP:ASC} \tag{LP:\,{Cover\,Small\,Cuts}}
\min \quad \qquad		& \quad \sum_{f \in L} c_f x_f 		& \\
\text{subject to: } 	& \quad \sum_{f\in L\cap\delta(S)} x_f \geq 1 	& \forall \, \, S \in \C \\
					& \quad 0\leq x_f \leq 1 & \forall \, \, f \in L. \\
\end{align*}

The first $O(1)$ approximation algorithm was presented by
Bansal et~al.~\cite{BCGI24}, and the approximation ratio was
improved from~16 to~10 by Nutov, \cite{N2024,N-waoa2024}, then from~10 to~6
by Bansal, \cite{B2023}, and later by Nutov, \cite{N2025}.

\begin{proposition} \label{prop:approxASC}
Given an instance of $\ASC$, the WGMV primal-dual algorithm, \cite{WGMV95}, finds
a feasible solution of cost $\leq \pliableapx\,\lpopt$ in polynomial time,
where $\lpopt$ denotes the optimal value of \eqref{LP:ASC}.
\end{proposition}
}

%%%%%
%	\subsubsection{The \texorpdfstring{$f$-connectivity}{f-connectivity} problem and Jain's iterative rounding algorithm}
\subsection{The {$f$-connectivity} problem and Jain's iterative rounding algorithm}
{
In the context of approximation algorithms, several connectivity
augmentation problems can be formulated in a general framework
called $f$-connectivity. In this problem, we are given an undirected
graph $G = (V,E)$ on $n$ nodes with nonnegative costs $c \in \Qp^E$
on the edges and a requirement function $f:2^V\to\Zp$ on subsets of nodes.
The algorithmic goal is to find an edge-set $\J \subseteq E$ with
minimum cost $c(\J) := \sum_{e \in \J} c_e$ such that for all cuts
$\delta(S),\ S \subseteq V$, we have $|\delta(S) \cap \J| \geq f(S)$.
{
A function $f$ is called \textit{weakly supermodular} if
$f(V)=0$, and
for all $A,B \subseteq{V}$, either
    $f(A) + f(B) \leq f(A - B) + f(B - A)$, or
    $f(A) + f(B) \leq f(A \cap B) + f(A \cup B)$.
}
The following is an LP~relaxation for the $f$-connectivity problem.

\begin{align*}\label{LP:f-conn} \tag{LP:\,{f-connectivity}}
\min \quad \qquad		& \quad \sum_{e \in E} c_e x_e 		& \\
\text{subject to: } 	& \quad \sum_{e\in E\cap\delta(S)} x_e \geq f(S) & \forall \; S \subset{V} \\
					& \quad 0\leq x_e \leq 1 & \forall \, \, e \in E. \\
\end{align*}

Assuming that the function $f$ is weakly supermodular, {integral,
and has a positive value for some $S\subset{V}$}, Jain \cite{Jain01}
presented a 2-approximation algorithm for the $f$-connectivity problem,
based on the following key result.

\begin{theorem}[Jain {\cite[Theorem~3.1]{Jain01}}]
Assume that the function $f$ is weakly supermodular, integral, non-negative and non-zero.
Then, in any basic solution $x$ to the LP relaxation,
for at least one edge, $e$, $x_e$ is at least $1/2$.
\end{theorem}
}

%%%%%
\subsection{Our results}
{
Our main result is the following.

\begin{theorem} \label{thm:main}
Let $k\geq4$ be an even integer.
There is an instance of the $\ASC$ problem $G=(V,E),L,u$
such that the LP relaxation \eqref{LP:ASC}
has a basic solution $x^*$ such that every positive variable of $x^*$ has value $1/k$.
\end{theorem}

In section~\ref{sec:capgraph}, we describe the capacitated graph $G=(V,E),u$
and the relevant family of small cuts.
In section~\ref{sec:linksgraph}, we describe the links graph $(V,L)$ and
the basic solution $x^*$.
In section~\ref{sec:A-rank}, we show that the basis matrix of $x^*$ has full rank.

Our theorem indicates that the polyhedron of feasible solutions to \eqref{LP:ASC}
differs in an essential way from the polyhedrons associated with several problems
in combinatorial optimization.
Moreover, our example shows that a direct application of Jain's iterative rounding algorithm
does not give an $O(1)$ approximation algorithm for $\ASC$.
We mention that Bansal~et~al.~\cite{BCGI24} were the first to present
an $O(1)$~approximation algorithm for $\ASC$. They showed that the
primal-dual method of Williamson~et~al.~\cite{WGMV95} achieves an
approximation ratio of $16$ for $\ASC$.
}
}
%%%%%%%%%%
{
\section{The Capacitated Graph \label{sec:capgraph}}

Let $k\ge4$ be an even integer.
The capacitated graph $G=(V,E),u$ has $n=2+\binom{k}{2}$ nodes,
and $m=(n-1)+(k-1)=\binom{k}{2}+k$ edges.
The edge capacities are chosen from the set $\{1,2,3\}$ and
$\lambda$ (the threshold value for small cuts) is chosen to be $5$.

We denote the nodes of $G$ by $v_1,v_2,\dots,v_n$, as well as by
the indices $1,2,\dots,n$.
For notational convenience, let $s=v_1$ and let $t=v_n$.

The family of small cuts consists of two sub-families.
One sub-family corresponds to the node-sets
$\N_1=\{v_1\}, 
 \N_2=\{v_1,v_2\}, \dots,
 \N_{n-1}=\{v_1,v_2,\dots,v_{n-1}\}$.
In other words, for each of the sets $\N_i=\{v_1,v_2,\dots,v_i\}, i=1,\dots,n-1$,
$\delta_E(\N_i)$ is a small cut.
We call the sets $\N_i$ ($i=1,\dots,n-1$) the \textit{nested sets}. 
The other sub-family corresponds to the $(k-1)$ node-sets
$\Q_1 = \{v_2,v_3,\dots,v_{1+\frac{k}{2}}\},
 \Q_2 = \{v_{2+\frac{k}{2}},v_{3+\frac{k}{2}},\dots,v_{1+2\cdot\frac{k}{2}}\}, \dots,
 \Q_{(k-1)} = \{v_{2+(k-2)\cdot\frac{k}{2}},v_{3+(k-2)\cdot\frac{k}{2}},\dots,v_{1+(k-1)\cdot\frac{k}{2}}\}$.
Note that these sets form a partition of $V(G)-\{s,t\}$ into $(k-1)$ sets, each of cardinality $k/2$.
In other words, for each of the $(k-1)$ ``intervals'' $\Q_j$ ($j=1,\dots,k-1$)
formed by picking the node with index $2+(j-1)\cdot\frac{k}{2}$
and the next $(\frac{k}{2}-1)$ consecutively indexed nodes,
$\delta_E(\Q_j)$ is a small cut.
We call the sets $\Q_j$ ($j=1,\dots,k-1$) the \textit{$\Q$-sets};
moreover, by the \textit{first~node} of $\Q_j$ we mean the node of lowest index in $\Q_j$
(i.e., node $2+(j-1)\cdot\frac{k}{2}$), and 
          by the \textit{last~node} of $\Q_j$ we mean the node of highest index in $\Q_j$
(i.e., node $1+(j)\cdot\frac{k}{2}$).
For notational convenience, 
let $\Q_0=\{s\}=\{v_1\}$ and let $\Q_k=\{t\}=\{v_n\}$;
moreover, let $s=v_1$ be the \textit{last~node} of $\Q_0$ and
          let $t=v_n$ be the \textit{first~node} of $\Q_k$.
(In the proofs below, if we examine the case that a specified node $v$
is the last~node of its $\Q$-set, then $v=s=v_1$ is possible,
whereas, if we examine the case that a specified node $v$
is the first~node of its $\Q$-set, then $v=s=v_1$ is not possible.
Similarly, 
if we examine the case that a specified node $v$
is the first~node (respectively, last~node) of its $\Q$-set,
then $v=t=v_n$ is possible (respectively, is not possible).)

For a nested set $\N_i$ ($i=1,\dots,n-1$) and a $\Q$-set $\Q_j$ ($j=1,\dots,k-1$),
observe that $s=v_1\in \N_i-\Q_j$ and $t=v_n\not\in \N_i\cup\Q_j$;
moreover, if $\N_i\cap\Q_j$ is non-empty, then either
$\Q_j$ crosses $\N_i$ (i.e., both $\Q_j-\N_i$ and $\Q_j\cap{\N_i}$ are non-empty) or
$\Q_j\subsetneq{\N_i}$.

There are two types of edges in $G$.
There are $n-1$ edges that form the $s,t$-path $v_1,v_2,\dots,v_n$;
we use $E_1$ to denote the edge~set of this path.
The first edge and the last edge of this path have capacity $2$,
thus, $u_{v_1v_2} = 2 = u_{v_{n-1}v_n}$.
Consider any other edge $v_iv_{i+1}$ $(i=2,\dots,n-2)$ of this path.
If both $v_i$ and $v_{i+1}$ are in the same $\Q$-set then $v_iv_{i+1}$ has capacity $3$,
otherwise, $v_iv_{i+1}$ has capacity one.
Moreover, there are $(k-1)$ other edges (that are not in the $s,t$-path),
and each of these edges has capacity one;
we use $E_2$ to denote this set of $(k-1)$ edges.
The nodes incident to the $j$-th edge of $E_2$ ($j=1,\dots,k-1$) have indices
$1+(j-1)\cdot\frac{k}{2}$ and $2+j\cdot\frac{k}{2}$.
Thus, $E_2$ has an edge between $s=v_1$ and the first~node of $\Q_2$,
an edge between the last~node of $\Q_{k-2}$ and $t=v_n$,
and, for each $j=1,\dots,k-3$, $E_2$ has an edge between the last~node of $\Q_j$
and the first~node of $\Q_{j+2}$.

The following figure illustrates the construction of the capacitated graph for $k=4$;
the graph has $\binom{k}{2}+2=8$ nodes and $\binom{k}{2}+k=10$ edges.

{
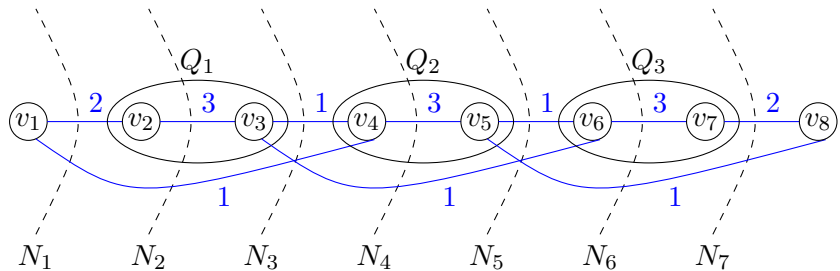
\begin{figure}[htb]
\begin{center}
\begin{tikzpicture}
    \foreach \i in {1,...,7} {
    \draw[dashed] (\i*1.5,-1.5) .. controls (\i*1.5+0.75,0) .. (\i*1.5,1.5);
    \node at (\i*1.5,-1.8) {$N_\i$};
    \node [draw,circle,minimum size=0.5cm,inner sep =0pt] at (\i*1.5-0.1,0) {$v_\i$};

    \draw[blue] (\i*1.5+0.15,0) -- (\i*1.5+1.15,0);
 }

 \foreach \j in {1,...,3} {
   \draw (\j*3+0.65,0) ellipse (1.2cm and 0.55cm);

    \node at (\j*3+0.65,0.8) {$Q_\j$};  

    \draw[blue] (\j*3-1.5,-0.23) .. controls (\j*3-0.25,-1.1) .. (\j*3+3,-0.24);
 }
 \node[draw,circle,minimum size=0.5cm,inner sep =0pt] at (8*1.5-0.1,0) {$v_8$};

\node at (2.3,0.25) {\bluetext{2}};
\node at (3.8,0.25) {\bluetext{3}};
\node at (5.3,0.25) {\bluetext{1}};
\node at (6.8,0.25) {\bluetext{3}};
\node at (8.3,0.25) {\bluetext{1}};
\node at (9.8,0.25) {\bluetext{3}};
\node at (11.3,0.25) {\bluetext{2}};
\node at (4,-1) {\bluetext{1}};
\node at (7,-1) {\bluetext{1}};
\node at (10,-1) {\bluetext{1}};
 
\end{tikzpicture}
\end{center}
\caption{ \label{fig:capacitated-graph}
The capacitated graph for $k=4$.
The small cuts are given by the nested~sets $\N_1,\dots,\N_7$ (inidicated by dashed lines)
and the sets $\Q_1,Q_2,\Q_3$ (indicated by ovals).}
\end{figure}
}

We defer the proofs of the next two results to the appendix;
our proofs are straightforward and use case analysis.

{
\begin{restatable}{proposition}{SmallCutsproposition}
\label{prop:smallcuts}
Each of the nested cuts $\delta(\N_i), i=1,\dots,n-1$
and each of the $\Q$-cuts $\delta(\Q_j), j=1,\dots,k-1$
has capacity less than $\lambda=5$.
\end{restatable}
}

{
\begin{restatable}{proposition}{BigCutsproposition}
\label{prop:bigcuts}
Every non-trivial cut $\delta(S)$, $\emptyset\subsetneq{S}\subsetneq{V}$, of $G$
that is neither a nested cut nor a $\Q$-cut
has capacity $\geq\lambda=5$.
\end{restatable}
}

}
%%%%%%%%%%
{
\section{The Links Graph \label{sec:linksgraph}}

Let $(V,L)$ denote the graph of the links;
thus $L$ is the set of links, and we denote the links by the symbol $\ell$, e.g., $\ell_i$, $\ell'$, etc.
The number of links is $m=n+k-2 = \binom{k}{2} + k$.
The links are partitioned into $k$ internally node disjoint $s,t$-paths
that we denote by $P_1,\dots,P_{k-1},P_k$.
The $s,t$-path $P_k$ has one link $st = v_1v_n$.
Each of the $s,t$-paths $P_1,\dots,P_{k-1}$ has $k/2$ internal nodes
and has $1+\frac{k}{2}$ links.
Moreover, the sequence of node indices of each of these $s,t$-paths is an
increasing sequence; that is, the node sequence has the form
$v_1,v_{i_2},v_{i_3},\dots,v_{i_{(\frac{k}{2}+1)}},v_n$, where
$1<i_2<i_3<\dots<i_{(\frac{k}{2}+1)}<n$.
Each of the nodes $v_2,\dots,v_{n-1}$ is in one of the $s,t$-paths $P_1,\dots,P_{k-1}$;
thus, each of these nodes is incident to two links.

Recall that we have $(k-1)$ $\Q$-sets, each with $k/2$ nodes,
and $(k-1)$ $s,t$-paths $P_1,\dots,P_{k-1}$ each with $k/2$ internal nodes.
Each of the $s,t$-paths $P_i$ $(i=1,\dots,k-1)$ is constructed such that
it is incident to precisely one node of $k/2$ (of the $(k-1)$) $\Q$-sets,
and it is disjoint from the the other $(k-1 - \frac{k}{2})$ $\Q$-sets.
The $s,t$-path $P_i$ $(i=1,\dots,k-1)$ is incident to the $\Q$-set with
index $i$ and the next $(k/2)-1$ consecutively indexed $\Q$-sets (with ``wrap around'');
thus, for ${i=1,\dots,k-1}$ and ${j=1,\dots,k-1}$, the $s,t$-path $P_i$ and the set $\Q_j$
have a node in common iff
\begin{align*}
j\geq{i} & \quad \textbf{and} \quad j \leq \min(i+\frac{k}{2}-1,\;k-1), \quad \textbf{or} \\
j<{i}    & \quad \textbf{and} \quad 1 \leq j \leq i-\frac{k}{2}.
\end{align*}
If an $s,t$-path $P_i$ and a set $\Q_j$ have a node in common,
then any one node of $\Q_j$ could be assigned to $P_i$;
in other words, our construction does not specify the internal nodes of $P_i$,
it only specifies the $\Q$-sets that intersect $P_i$.

Let $A^{PQ}$ denote the $(k-1)\times(k-1)$ zero-one matrix that has
a row for each $s,t$-path $P_i$ $(i=1,\dots,k-1)$ and
a column for each set $\Q_j$ $(i=1,\dots,k-1)$ such that
the entry $A^{PQ}_{i,j}$ is one iff $P_i$ and $\Q_j$ have a node in common.
Then, by construction, $A^{PQ}$ is a circulant matrix with entries of zero or one
such that each row has $k/2$ consecutive ones (with ``wrap around'')
and the remaining $(k-1-\frac{k}{2})$ entries of the row are zero.
Observe that the number of ones in each row, $k/2$, and the size of the matrix, $k-1$,
are relatively prime (i.e., $\gcd(\frac{k}{2},k-1)=1$).
For an illustration, see the example at the end of section~\ref{sec:A-rank}.

\begin{lemma}	\label{lem:circulant-rank}
The matrix $A^{PQ}$ has rank $(k-1)$. Moreover, $\det(A^{PQ})=k/2$.
\end{lemma}

One can prove this lemma directly, by applying row operations
(and applying the matrix determinant lemma to a square matrix
with diagonal entries $2$ and all other entries one).
% \url{https://en.wikipedia.org/wiki/Matrix_determinant_lemma}
Alternatively, the lemma follows directly from a result of Hariprasad \cite[Theorem~2.1]{Hari2019}.

Figure~\ref{fig:links-graph} illustrates the construction of the links graph for $k=4$.
The graph has $\binom{k}{2}+2=8$ nodes and $\binom{k}{2}+k=10$ links.
The links partition into $k=4$ (internally node disjoint) $s,t$-paths $P_1,\dots,P_4$;
the links of $P_1,\dots,P_4$ are indicated by distinct colours.

{
\begin{figure}[htb]
\begin{center}
\begin{tikzpicture}
    \foreach \i in {1,...,7} {
    \draw[dashed] (\i*1.5,-1.5) .. controls (\i*1.5+0.75,0) .. (\i*1.5,1.5);
    \node at (\i*1.5,-1.8) {$N_\i$};
    \node [draw,circle,minimum size=0.5cm,inner sep =0pt] at (\i*1.5-0.1,0) {$v_\i$};
 }

 \foreach \j in {1,...,3} {
   \draw (\j*3+0.65,0) ellipse (1.2cm and 0.55cm);
    \node at (\j*3+0.65,0.8) {$Q_\j$};  
 }
 \node[draw,circle,minimum size=0.5cm,inner sep =0pt] at (8*1.5-0.1,0) {$v_8$};

\draw[blue] (1.65,0) -- (2.65,0);
\draw[blue] (3,-0.22) .. controls (4.5,-0.8) .. (5.8,-0.24);
\draw[blue] (6,-0.22) .. controls (9.5,-1) .. (11.65,-0.1);

\draw[red] (1.5,0.22) .. controls (4.5,1.5) .. (7.3,0.25);
\draw[red] (7.65,0) -- (8.65,0);
\draw[red] (8.9,0.25) .. controls (9.5,1.5) .. (11.8,0.24);

\draw[teal] (1.45,-0.22) .. controls (2.9
,-0.4) .. (4.4,-0.24);
\draw[teal] (4.4,0.24) .. controls (7,1.5) .. (10.3,0.24);
\draw[teal] (10.65,0) -- (11.65,0);

\draw[brown] (1.5,-0.22) .. controls (8,-1.8) .. (11.8,-0.24);

\node at (2.3,0.2) {\bluetext{$\ell_1$}};
\node at (3.8,-0.1) {\tealtext{$\ell_3$}};
\node at (4.1,1.4) {\redtext{$\ell_2$}};
\node at (7,-1.2) {\browntext{$\ell_4$}};
\node at (5.3,-0.7) {\bluetext{$\ell_5$}};
\node at (7,1.4) {\tealtext{$\ell_6$}};
\node at (8.2,-0.9) {\bluetext{$\ell_7$}};
\node at (8.3,0.2) {\redtext{$\ell_8$}};
\node at (10.1,1.3) {\redtext{$\ell_9$}};
\node at (11.3,0.2) {\tealtext{$\ell_{10}$}};
 
\end{tikzpicture}
\end{center}
\caption{ \label{fig:links-graph}
The links graph for $k=4$.
The links partition into $s,t$-paths $P_1,\dots,P_4$, and
the links of each of these $s,t$-paths is indicated by a distinct colour.
Thus, the links of $P_1$ are $\ell_1,\ell_5,\ell_7$,
      the links of $P_2$ are $\ell_2,\ell_8,\ell_9$,
      the links of $P_3$ are $\ell_3,\ell_6,\ell_{10}$, and
      $P_4$ has the link $\ell_4$.
The small cuts are given by the nested~sets $\N_1,\dots,\N_7$ (inidicated by dashed lines)
and the sets $\Q_1,Q_2,\Q_3$ (indicated by ovals).}
\end{figure}
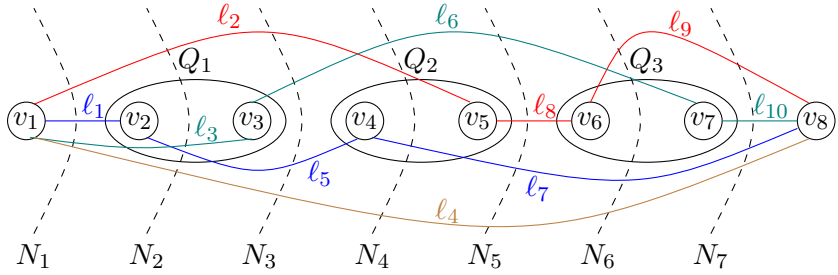
}

We construct a basic feasible solution $x^*$ by assigning the value $1/k$ to each link,
thus, $x^*_{\ell}=1/k, \forall\ell\in{L}$.
Informally speaking, we are assigning a value $1/k$ to each of the $s,t$-paths $P_1,\dots,P_k$.
It is easy to verify that each of the small cuts,
namely, $\delta(\N_i)$ $(i=1,\dots,n-1)$ and $\delta(\Q_j)$ $(j=1,\dots,k-1)$,
is ``tightly covered'' by $x^*$.
First, consider a cut $\delta(\N_i)$ $(i=1,\dots,n-1)$,
and observe that each of the $s,t$-paths $P_1,\dots,P_k$
has exactly one of its links in $\delta(\N_i)$, hence, $x^*(\delta(\N_i))=(k)\frac{1}{k}=1$.
Next, consider a cut $\delta(\Q_j)$ $(j=1,\dots,k-1)$,
and observe that $k/2$ of the $s,t$-paths $P_1,\dots,P_k$
have exactly one internal node in $\Q_j$,
hence, each of these $s,t$-paths
has exactly two links in $\delta(\Q_j)$, therefore,
	$x^*(\delta(\Q_j))=2(\frac{k}{2})(\frac{1}{k})=1$.

The basis matrix $A$ of $x^*$ has one row for each small cut and
one column for each link;
the entry of $A$ corresponding to a small cut $\delta(S)$ and a link $\ell$
is one if $\ell\in\delta(S)$, otherwise, the entry is zero.
Thus, $A$ is an $m\times{m}$ matrix with entries of zero or one;
recall that $m = n+k-2 = \binom{k}{2}+k$.
In the next section, we show that matrix $A$ has full rank,
and that will prove our claim that $x^*$ is a basic feasible solution
of the LP~relaxation of $\ASC$.

}
%%%%%%%%%%
{
\section{The rank of the incidence matrix of small cuts and links \label{sec:A-rank}}

Let $A$ denote the zero-one incidence matrix of the small cuts and the links;
$A$ is an $m\times{m}$ matrix, where $m = n+k-2 = \binom{k}{2}+k$.
The matrix $A$ has rows corresponding to the cuts of the $\Q$-sets
followed by rows corresponding to the cuts of the nested sets;
the columns of $A$ correspond to the links.
In this section, the main goal is to show that $\det(A)$ is non-zero;
equivalently, $A$ has full rank.
We start by fixing a convenient indexing for the rows and columns of $A$.
The rows $1,\dots,(k-1)$ of $A$ correspond to the cuts of $\Q_1,\dots,\Q_{(k-1)}$,
and the rows $k,(k+1),\dots,m$ of $A$ correspond to the cuts of $\N_1,\dots,\N_{(n-1)}$.
The links $\ell\in{L}$ are indexed as follows:
The link of the $s,t$-path $P_i$ $(i=1,\dots,k)$ incident to node $s=v_1$ is denoted $\ell_i$;
thus, each of $\ell_1,\dots,\ell_k$ is in $\delta(s)$, and there is a bijection between
these links and the $s,t$-path $P_1,\dots,P_k$.
Note that each of the remaining links is incident to one of the nodes $v_2,\dots,v_{n-1}$,
and, for each $i=2,\dots,(n-1)$,
the unique link that is incident to node $v_i$ and is in the cut $\delta(\N_i)$ is assigned
the index $(k+i-1)$.
The columns $1,\dots,m$ of $A$ correspond to the links $\ell_1,\dots,\ell_m$.

Let us view the matrix $A$ as a $2\times{2}$ block matrix
$\begin{pmatrix}
A_{11} & A_{12} \\
A_{21} & A_{22}
\end{pmatrix}$
where $A_{11}$ is the $(k-1)\times(k-1)$ sub-matrix of $A$
whose rows correspond to the cuts of $\Q_1,\dots,\Q_{(k-1)}$ and
whose columns correspond to the links $\ell_1,\dots,\ell_{(k-1)}$,
and
$A_{22}$ is the $(m-k+1)\times(m-k+1)$ sub-matrix of $A$
whose rows correspond to the cuts of $\N_1,\dots,\N_{(n-1)}$ and
whose columns correspond to the links $\ell_k,\ell_{(k+1)},\dots,\ell_{m}$.

\begin{claim}
The sub-matrix $A_{22}$ is lower-triangular and all entries on the diagonal are one;
thus, $\det(A_{22})=1$.
\end{claim}
\begin{proof}
The link $\ell_k$ (with end~nodes $s$ and $t$) is in each of the
cuts $\delta(\N_i)$, $i=1,\dots,(n-1)$.
For each of the links $\ell_{(k+i-1)}$ where $i=2,\dots,(n-1)$, note that
the link is in the cut $\delta(\N_i)$ and it is in none of the cuts
$\delta(\N_1),\dots,\delta(N_{(i-1)})$.
\end{proof}

Our plan is to apply row operations to the first $(k-1)$ rows of $A$
in order to transform the submatrix
$\begin{pmatrix}
A_{11} & A_{12} \\
\end{pmatrix}$
to a submatrix of the form
$\begin{pmatrix}
(A^{PQ})^{\top} & \textbf{0}_{(k-1)\times(m-k+1)}\\
\end{pmatrix},$
where $(A^{PQ})^{\top}$ denotes the transpose of the matrix $A^{PQ}$ and
$\textbf{0}_{(k-1)\times(m-k+1)}$ denotes a ${(k-1)\times(m-k+1)}$ matrix of zeros.
Therefore, using row operations, we will rewrite the matrix $A$ as
the $2\times{2}$ block matrix
$\begin{pmatrix}
(A^{PQ})^{\top} & \textbf{0}_{(k-1)\times(m-k+1)}\\
A_{21} & A_{22} \\
\end{pmatrix};$
then, by applying Lemma~\ref{lem:circulant-rank} as well as a result on
the determinant of $2\times{2}$ block matrices,
we have $\det(A) = \det(A^{PQ}) \det(A_{22}) = \det(A^{PQ}) \not=0$.

See the example at the end of this section for an illustration.

Let us introduce some notation to describe and analyse our row operations.
For any set of links $L'$ let $\chi^{L'}$ denote the zero-one row incidence vector
of $L'$; thus, for any link $\ell$, $\chi^{L'}_{\ell}=1$ iff $\ell\in{L'}$.
Note that rows $1,\dots,(k-1)$ of $A$ correspond to
$\chi^{\delta(\Q_1)},\dots,\chi^{\delta(\Q_{(k-1)})}$.
For any link $\ell$, let $\p(\ell)$ denote the index $i$ of the $s,t$-path $P_i$
that contains $\ell$.
For each of the links $\ell_i$, $i=1,\dots,(k-1)$, note that $\p(\ell_i) = i$,
since we have a bijection between these links and the $s,t$-paths $P_1,\dots,P_{(k-1)}$.
For any set of links $L'$, let $\hp(L')$ denote the set of indices of
the $s,t$-paths $P_1,\dots,P_{(k-1)}$ that contain at least one of the links of $L'$;
thus, $\hp(L')\subseteq\{1,\dots,k\}$, and
index $i$ is in $\hp(L')$ iff the $s,t$-path $P_i$ contains one of the links of $L'$.
For example, if $L'$ is a subset of $\{\ell_1,\dots,\ell_{(k-1)}\}$,
then $\hp(L') = \{ i \;:\; \ell_i\in{L'} \}$.

Consider any row $j$, $j=1,\dots,(k-1)$ of the matrix $A$;
this row-vector is the same as $\chi^{\delta(\Q_j)}$.
Let $g$ denote the largest index $i$ such that the nested set $\N_i$
is disjoint from $\Q_j$, and let $h$ denote the smallest index $i$
such that the nested set $\N_i$ contains $\Q_j$.
(By our construction, the $k/2$ nodes of $\Q_j$ are in the interval
$[(g+1),(g+2),\dots,h]$.)

\begin{claim} \label{cl:Q-operations}
$\ds \big( \chi^{\delta(\Q_j)} - \chi^{\delta(\N_h)} + \chi^{\delta(\N_g)} \big)
\:=\: 2\chi^{L'},$
where $L'$ is the set of $k/2$ links in $\delta(\Q_j) \cap \delta(\N_g)$.
\end{claim}
\begin{proof}
By construction, every link has its two end~nodes in two different $\Q$-sets.
Hence, every link $\ell\in\delta(\Q_j)$ that has its lower-index node in $\Q_j$
is in $\delta(\N_h)$; similarly,
every link $\ell\in\delta(\Q_j)$ that has its higher-index node in $\Q_j$
is in $\delta(\N_g)$.
Moreover, $\delta(\N_g) - \delta(\Q_j) = \delta(\N_h) - \delta(\Q_j)$
is the set of links that have their lower-index node in $\N_g$
and their higher-index node in $V-\N_h$.
The equation in the claim follows from these statements.
\end{proof}

In the above claim, note that $L'$ is contained in the cut of a $\Q$-set, hence,
the link $\ell_k=st$ cannot be in $L'$.

Our plan is to (separately) apply the row operations described in the above claim
to each of the rows $j=1,\dots,(k-1)$ of $A$.
The next lemma explains how to apply row operations to $\chi^{L'}$,
where $L'$ is any set of $k/2$ links contained in some nested cut $\delta(\N_i)$
and $\ell_k\not\in{L'}$,
to transform $\chi^{L'}$ to the vector $\chi^{\hL}$,
where $\hL = \{ \ell_i \;:\; i\in\hp(L') \}$.
Informally speaking, we can apply row operations to ``map'' any row vector $\chi^{L'}$,
where $L'\subset\delta(\N_i)-\{\ell_k\}$ and $|L'|=k/2$,
to the row incidence vector (padded with zeros on the right)
of the $k/2$ $s,t$-paths among $P_1,\dots,P_{(k-1)}$
that contain (at least) one of the links of $L'$.
Thus, by applying the above claim and the next lemma,
we ``map'' each row of the submatrix
$\begin{pmatrix}
A_{11} & A_{12} \\
\end{pmatrix}$
to a distinct row of the submatrix
$\begin{pmatrix}
(A^{PQ})^{\top} & \textbf{0}_{(k-1)\times(m-k+1)}\\
\end{pmatrix}.$

\begin{lemma} \label{lem:move-linkset}
Let $i=1,\dots,n-1$, and let $L'\subset\delta(\N_i)-\{\ell_k\}$ be any set of $k/2$ links.
By applying row operations, one can transform the vector $\chi^{L'}$
to the vector $\chi^{\hL}$,
where $\hL = \{ \ell_i \;:\; i\in\hp(L') \}$.
\end{lemma}
\begin{proof}
By induction on the index of the nested set whose cut contains $L'$.

Let $h$ be the smallest index such that the nested cut $\delta(\N_h)$ contains $L'$.
If $h=1$, then we have $\hL=L'$ and the lemma is proved.

Now, suppose $h\geq2$; thus, $L'$ contains a link $\ell_j$ with $j>k$.
Observe that $L'$ contains the link $\ell_{(h+k-1)}$ which is incident to node $v_h$
and is in the cut $\delta(\N_h)$;
informally speaking, the presence of $\ell_{(h+k-1)}$ in $L'$ certifies that
$L'$ is contained in $\delta(\N_h)$ and $L'$ is not a subset of $\delta(\N_j)$ for any $j<h$.
Let $\bL$ denote the set of links $\delta(\N_h)-L'$;
$\bL$ is a ``complementary set'' of $k/2$ links w.r.t.\ $\delta(\N_h)$ and $L'$.
For every link $\ell_j\in\bL$, observe that $j<(h+k-1)$
(since $\ell_j\in\delta(N_h)$ and $\ell_{(h+k-1)}\not\in\bL$).

Let $g$ be the smallest index such that the nested cut $\delta(\N_g)$ contains $\bL$.
Again, observe that $\bL$ contains the link $\ell_{(g+k-1)}$ which is incident to node $v_g$
and is in the cut $\delta(\N_g)$.
Clearly, $g<h$ and $\delta(\N_g)\not=\delta(\N_h)$,
since $\ell_{(h+k-1)}\in\delta(\N_h)$ and $\ell_{(h+k-1)}\not\in\delta(\N_g)$.

\begin{claim} \label{cl:move-operations}
There exists a set of $k/2$ links of $\delta(\N_g)-\{\ell_k\}$, call it $L''$, such that
\\
(i)\quad $\ds \chi^{L''} = \chi^{L'} - \chi^{\delta(\N_h)} + \chi^{\delta(\N_g)}$, and
\\
(ii)\quad $\ds \hp(L'') = \hp(L')$.
\end{claim}
\begin{proof}
Observe that $\bL = \delta(\N_h)-L'$, and that $\bL$ is contained in $\delta(\N_g)$;
also, note that $\ell_k$ is in $\bL$.
Let $L'' = \delta(\N_g) - \bL$;
observe that $L''$ is a set of $k/2$ links and $\ell_k\not\in{L''}$.
Hence, we have $\chi^{\delta(\N_g)} - \chi^{\delta(\N_h)} = \chi^{L''} - \chi^{L'}$.
This proves part~(i) of the claim.

Moreover, note that each of the nested cuts contains exactly one link from each of the
$s,t$-paths $P_1,\dots,P_k$;
that is, for each $i=1,\dots,n-1$, $\hp(\delta(\N_i)) = \{1,\dots,k\}$.
Hence,
$\ds \hp(L') = \hp(\delta(\N_h)-\bL) = \{1,\dots,k\}-\hp(\bL) = \hp(\delta(\N_g)-\bL)
	= \hp(L'')$.
This proves part~(ii) of the claim.
\end{proof}

In the induction step (of the proof of the lemma), we apply the above claim
to replace the set $L'$ by the set $L''$;
note that $L''$ has the properties of $L'$ (listed in the lemma statement) and
$\hp(L'') = \hp(L')$.
Moreover, $L''$ is contained in the cut of a nested set of smaller index
than the index of any nested set whose cut contains $L'$.

Thus, the lemma follows by induction.
\end{proof}

\subsection*{Example with $k=4$}
{
In this subsection, we fix $k=4$, and we illustrate our row operations,
in particular, the row operations of Claims~\ref{cl:Q-operations},\ref{cl:move-operations},
on the links graph shown in Figure~\ref{fig:links-graph}.
The links partition into $s,t$-paths $P_1,\dots,P_4$, and
      the links of $P_1$ are $\ell_1,\ell_5,\ell_7$,
      the links of $P_2$ are $\ell_2,\ell_8,\ell_9$,
      the links of $P_3$ are $\ell_3,\ell_6,\ell_{10}$, and
      $P_4$ has the link $\ell_4$.

The incidence matrix $A^{PQ}$ of the $s,t$-paths $P_1,P_2,P_3$
and the $\Q$-sets is given below.
\[
\begin{blockarray}{cccc}
Q_1 & Q_2 & Q_3 \\
\begin{block}{(ccc)c}
  1 & 1 & 0 & P_1 \\
  0 & 1 & 1 & P_2 \\
  1 & 0 & 1 & P_3 \\
\end{block}
\end{blockarray}
 \]

The incidence matrix $A$ of the small cuts and the links is given below.
Note that the submatrix $A_{22}$ (of the nested cuts and the links $\ell_4,\dots,\ell_{10}$)
is lower-triangular with diagonal entries of one.
\[
\begin{blockarray}{ccc||cccccccc}
\ell_1 & \ell_2 & \ell_3 & \ell_4 & \ell_5 & \ell_6 & \ell_7 & \ell_8 & \ell_9 & \ell_{10} \\
\begin{block}{(ccc||ccccccc)c}
  1 & 0 & 1 & 0 & 1 & 1 & 0 & 0 & 0 & 0 & Q_1 \\
  0 & 1 & 0 & 0 & 1 & 0 & 1 & 1 & 0 & 0 & Q_2 \\
  0 & 0 & 0 & 0 & 0 & 1 & 0 & 1 & 1 & 1 & Q_3 \\
\BAhline\BAhline
  1 & 1 & 1 & \circled{1} & 0 & 0 & 0 & 0 & 0 & 0 & N_1 \\
  0 & 1 & 1 & 1 & \circled{1} & 0 & 0 & 0 & 0 & 0 & N_2 \\
  0 & 1 & 0 & 1 & 1 & \circled{1} & 0 & 0 & 0 & 0 & N_3 \\
  0 & 1 & 0 & 1 & 0 & 1 & \circled{1} & 0 & 0 & 0 & N_4 \\
  0 & 0 & 0 & 1 & 0 & 1 & 1 & \circled{1} & 0 & 0 & N_5 \\
  0 & 0 & 0 & 1 & 0 & 1 & 1 & 0 & \circled{1} & 0 & N_6 \\
  0 & 0 & 0 & 1 & 0 & 0 & 1 & 0 & 1 & \circled{1} & N_7 \\
\end{block}
\end{blockarray}
 \]

We describe the row operations for $\delta(\Q_1), \delta(\Q_2), \delta(\Q_3)$,
i.e., rows 1,2,3 of $A$;
in particular, we describe the row operations of
Claims~\ref{cl:Q-operations},\ref{cl:move-operations}.

\begin{description}
\item[$\Q_1$:]\quad
$\chi^{\delta(\Q_1)} = \chi^{\{\ell_1,\ell_3,\ell_5,\ell_6\}}$.
In Claim~\ref{cl:Q-operations}, for $j=1$, we have $h=3, g=1$.
$\ds \big( \chi^{\delta(\Q_1)} - \chi^{\delta(\N_3)} + \chi^{\delta(\N_1)} \big)
\:=\: 2\chi^{\{\ell_1,\ell_3\}}.$

Note that $\hp(\delta(\Q_1)) = \{1,3\}$.

\item[$\Q_2$:]\quad
$\chi^{\delta(\Q_2)} = \chi^{\{\ell_2,\ell_5,\ell_7,\ell_8\}}$.
In Claim~\ref{cl:Q-operations}, for $j=2$, we have $h=5, g=3$.
$\ds \big( \chi^{\delta(\Q_2)} - \chi^{\delta(\N_5)} + \chi^{\delta(\N_3)} \big)
\:=\: 2\chi^{\{\ell_2,\ell_5\}}.$

Next, Claim~\ref{cl:move-operations} is applied with
$L' = \{\ell_2,\ell_5\}$, and we have $h=2, g=1$.
$\ds \chi^{\{\ell_2,\ell_5\}} - \chi^{\delta(\N_2)} + \chi^{\delta(\N_1)} =
	\chi^{\{\ell_1,\ell_2\}}$.

Note that $\hp(\delta(\Q_2)) = \{1,2\}$.

\item[$\Q_3$:]\quad
$\chi^{\delta(\Q_3)} = \chi^{\{\ell_6,\ell_8,\ell_9,\ell_{10}\}}$.
In Claim~\ref{cl:Q-operations}, for $j=3$, we have $h=7, g=5$.
$\ds \big( \chi^{\delta(\Q_3)} - \chi^{\delta(\N_7)} + \chi^{\delta(\N_5)} \big)
\:=\: 2\chi^{\{\ell_6,\ell_8\}}.$

Next, Claim~\ref{cl:move-operations} is applied with
$L' = \{\ell_6,\ell_8\}$, and we have $h=5, g=4$.
$\ds \chi^{\{\ell_6,\ell_8\}} - \chi^{\delta(\N_5)} + \chi^{\delta(\N_4)} =
	\chi^{\{\ell_2,\ell_6\}}$.

Next, Claim~\ref{cl:move-operations} is applied with
$L' = \{\ell_2,\ell_6\}$, and we have $h=3, g=2$.
$\ds \chi^{\{\ell_2,\ell_6\}} - \chi^{\delta(\N_3)} + \chi^{\delta(\N_2)} =
	\chi^{\{\ell_2,\ell_3\}}$.

Note that $\hp(\delta(\Q_3)) = \{2,3\}$.

\end{description}

}

}
%%%%%%%%%%
\bibliographystyle{plainurl}
\bibliography{example-IRT}
%%%%%%%%%%
% APPENDIX %
%%%%%%%%%%
\clearpage

\begin{appendices}
%%%%%%%%%%
\section{Deferred Proofs \label{sec:A1:proofs}}
{       
In this appendix, we present some deferred proofs.

%%%%%
{
\SmallCutsproposition*

\IGNORE{
\begin{restatable}{proposition}{SmallCutsproposition}
\label{prop:smallcuts}
Each of the nested cuts $\delta(\N_i), i=1,\dots,n-1$
and each of the $\Q$-cuts $\delta(\Q_j), j=1,\dots,k-1$
has capacity less than $\lambda=5$.
\end{restatable}
}

\begin{proof}
We examine a few cases.

\begin{description}

\item\textbf{Nested cut $\delta(\N_i)$ such that $\N_i$ does not cross any $\Q$-set:}

Then, either 
$2\leq i\leq n-2$ and there is a largest index $j=1,\dots,k-2$
such that $\Q_j$ is contained in $\N_i$,
or else $i\in\{1,\;(n-1)\}$.

If $i=1$, then $\delta(\N_i)$ has two edges, namely,
(i)~the $E_1$-edge $v_1v_{2}$ of capacity $2$,
and
(ii)~the $E_2$-edge (of capacity one) between $s=v_1$ and the first~node of $\Q_{2}$;
clearly, $\delta(\N_1)$ has capacity $3 <\lambda=5$.
Similarly, if $i=(n-1)$, then $\delta(\N_i)$ has two edges and has capacity $3 <\lambda=5$.

Otherwise, $\delta(\N_i)$ has three edges, namely,
(i)~the $E_1$-edge $v_iv_{i+1}$ of capacity one,
(ii)~the $E_2$-edge (of capacity one)
between the last~node of $\Q_{j-1}$ and the first~node of $\Q_{j+1}$,
and
(iii)~the $E_2$-edge (of capacity one)
between the last~node of $\Q_{j}$ and the first~node of $\Q_{j+2}$,
Thus, the cut $\delta(\N_i)$ has capacity $3 <\lambda=5$.

\item\textbf{Nested cut $\delta(\N_i)$ such that $\N_i$ crosses a $\Q$-set:}

Suppose that $\N_i$ crosses $\Q_j$, where $j=1,\dots,k-1$.
Clearly, the first~node of $\Q_j$ is in $\N_i$ and the last~node of $\Q_j$ is not in $\N_i$.
Note that $\delta(\N_i)$ has two edges, namely,
the $E_1$-edge $v_iv_{i+1}$ of capacity $3$,
and the $E_2$-edge (of capacity one)
between the last~node of $\Q_{j-1}$
and the first~node of $\Q_{j+1}$.
Thus, the cut $\delta(\N_i)$ has capacity $4 < \lambda=5$.

\item\textbf{$\Q$-cut $\delta(\Q_j)$:}

First, supose that $j=2,\dots,(k-2)$.
Observe that $\delta(\Q_j)$ consists of the following four unit-capacity edges.
There are two edges of $E_1$ in the cut,
namely, the edge between the last~node of $\Q_{j-1}$
and the first~node of $\Q_j$, and
        the edge between the last~node of $\Q_{j}$
and the first~node of $\Q_{j+1}$
Moreover, there are two edges of $E_2$ in the cut;
namely, the edge between the last~node of $\Q_{j-2}$
and the first~node of $\Q_j$, and
        the edge between the last~node of $\Q_{j}$
and the first~node of $\Q_{j+2}$.

The cuts $\delta(\Q_1)$ and $\delta(\Q_{k-1})$ have capacity four
since each of these cuts 
has only one edge of $E_2$
(since no $E_2$-edge is incident to the first~node of $\Q_1$ or to the last~node of $\Q_{k-1}$),
and
has two edges of $E_1$,
one edge of capacity $2$ (i.e., $sv_2\in\delta(\Q_1)$ and $v_{n-1}t\in\delta(\Q_{k-1})$)
and the other edge of capacity one.

\end{description}
\end{proof}
}

%%%%%
{
\BigCutsproposition*

\IGNORE{
\begin{restatable}{proposition}{BigCutsproposition}
\label{prop:bigcuts}
Every non-trivial cut $\delta(S)$, $\emptyset\subsetneq{S}\subsetneq{V}$, of $G$
that is neither a nested cut nor an $\Q$-cut
has capacity $\geq\lambda=5$.
\end{restatable}
}

\begin{proof}
We denote the nodes $v_1,\dots,v_n$ by their indices $1,\dots,n$ for this proof.
W.l.o.g., assume that $S$ is a set of nodes such that $1\in S$ (thus, $S$ contains $v_1$),
$S\neq\N_i, i=1,\dots,n-1$, and $S\neq(V-\Q_j), j=1,\dots,k-1$.
Let $\ell_1=1$ and let $r_1$ denote the smallest index such that $r_1\in{S}$
and $(r_1+1)\not\in{S}$;
note that $r_1$ is well~defined, since $S\neq{V}$; possibly, $r_1=\ell_1$.
Thus, $[\ell_1,\dots,r_1]$ is a maximal interval contained in $S$.
Clearly, $S\not= [\ell_1,\dots,r_1]$, otherwise, we would have $S=\N_{r_1}$
(contradicting our assumption that $S$ is not a nested set).
Thus, $r_1\leq(n-2)$.
Let $\ell_2$ be the smallest index $> r_1$ that is in $S$;
thus, $\ell_2 = \min_i \{i\in{S}, i \in [(r_1+1),\dots,n]\}$.
Moreover, let $r_2$ be the smallest index $\geq \ell_2$ such that
either $r_2=n$ or else $(r_2+1)\not\in{S}$;
thus, $r_2 = n$ if $[\ell_2,\dots,n]\subset{S}$, and
$r_2 = \min\{ i\in{S}, i+1\not\in{S}, i \in [\ell_2,\dots,n] \}$ if $[\ell_2,\dots,n]\not\subset{S}$.
Then, $[\ell_2,\dots,r_2]$ is another maximal interval contained in $S$.
Clearly, both the $E_1$-edges $v_{r_1}v_{(r_1+1)}$ and $v_{(\ell_2-1)}v_{\ell_2}$ are in $\delta(S)$.

For any node $i$ ($i=1,\dots,n$), we use $\q(i)$ to denote the index of the (unique) $\Q$-set
that contains $i$; thus, $\q(i) = \lceil (i-1)/\frac{k}{2} \rceil$.

We examine a few cases and show that $\delta(S)$ has capacity $\geq\lambda=5$ in each case.

\begin{description}

\item\textbf{Node $r_1$ is not the last~node of its $\Q$-set, $\Q_{\q(r_1)}$, and
      node $\ell_2$ is not the first~node of its $\Q$-set, $\Q_{\q(\ell_2)}$:}

Note that $r_1\not=s=1$ and $\ell_2\not={t}=n$.

Then the $E_1$-edge $v_{r_1}v_{(r_1+1)}$ has capacity $3$, and
the $E_1$-edge $v_{(\ell_2-1)}v_{\ell_2}$ has capacity $3$,
hence, $\delta(S)$ has capacity $\geq6 >\lambda=5$.

\item\textbf{Node $r_1$ is not the last~node of its $\Q$-set, $\Q_{\q(r_1)}$, and
      node $\ell_2$ is the first~node of its $\Q$-set, $\Q_{\q(\ell_2)}$:}

Note that $r_1\not=s=1$, and, possibly, $\ell_2={t}=n$.

Thus, $r_1\geq2$.
Then the $E_1$-edge $v_{r_1}v_{(r_1+1)}$ has capacity $3$.

Suppose $\ell_2={t}=n$.
Then the $E_1$-edge $v_{(\ell_2-1)}v_{\ell_2}$ has capacity two.
Thus, $\delta(S)$ has capacity $\geq5 =\lambda=5$.

Now, suppose $\ell_2\not={t}=n$.
Then the $E_1$-edge $v_{(\ell_2-1)}v_{\ell_2}$ has capacity one.
Note that $\q(r_1)\geq1$ and $2\leq\q(\ell_2)\leq(k-1)$.
We have two subcases, depending on $\q(\ell_2) - \q(r_1)$.

\begin{description}

\item[$\q(\ell_2) = \q(r_1)+1$:]
Then we have two further subcases.
\\
\textbf{(i)} 
First, suppose that $r_2=n$;
thus, $[\ell_2,\dots,n]$ is a maximal interval contained in $S$.
Then the last node of $\Q_{\q(r_1)}$ is not in $S$ and the first node of $\Q_{(\q(r_1)+2)}$ is in $S$.
Then the $E_2$-edge between the last node of $\Q_{\q(r_1)}$ and the first node $\Q_{(\q(r_1)+2)}$ is in $\delta(S)$.
Thus, the cut $\delta(S)$ has capacity $\geq5 = \lambda=5$.
\\
\textbf{(ii)} 
Now, suppose that $r_2<n$.
Then the $E_1$-edge $v_{r_2}v_{(r_2+1)}$ is in $\delta(S)$, and it has capacity one or more.
Thus, the cut $\delta(S)$ has capacity $\geq5 = \lambda=5$.

\item[$\q(\ell_2) \geq \q(r_1)+2$:]
Then the last node of $\Q_{(\q(r_1)-1)}$ is in $S$ and the first node of $\Q_{(\q(r_1)+1)}$ is not in $S$.
Then the $E_2$-edge between the last node of $\Q_{(\q(r_1)-1)}$ and the first node $\Q_{(\q(r_1)+1)}$ is in $\delta(S)$.
Thus, the cut $\delta(S)$ has capacity $\geq5 = \lambda=5$.

\end{description}

\item\textbf{Node $r_1$ is the last~node of its $\Q$-set, $\Q_{\q(r_1)}$, and
      node $\ell_2$ is not the first~node of its $\Q$-set, $\Q_{\q(\ell_2)}$:}

Possibly, $r_1=s=1$, and recall that $r_1\leq(n-2)$.
Note that $\ell_2\not={t}=n$.

	First, suppose $r_1=1$. 
	Then the $E_1$-edge $v_{r_1}v_{(r_1+1)}$ has capacity two,
	and the $E_1$-edge $v_{(\ell_2-1)}v_{\ell_2}$ has capacity $3$.
	Hence, $\delta(S)$ has capacity $\geq5 =\lambda=5$.

Now, suppose $r_1\geq2$. Then, $1\leq\q(r_1)\leq(k-2)$.
Then the $E_1$-edge $v_{r_1}v_{(r_1+1)}$ has capacity one,
and the $E_1$-edge $v_{(\ell_2-1)}v_{\ell_2}$ has capacity $3$.
Clearly, $\delta(S)$ has both these edges, as well as the $E_2$-edge
between the last~node of $\Q_{\q(r_1)-1}$ and the node $(r_1+1)$ (first~node of $\Q_{\q(r_1)+1}$).
 Hence, $\delta(S)$ has capacity $\geq5 =\lambda=5$.

\item\textbf{Node $r_1$ is the last~node of its $\Q$-set, $\Q_{\q(r_1)}$, and
      node $\ell_2$ is the first~node of its $\Q$-set, $\Q_{\q(\ell_2)}$:}
{
Possibly, $r_1=s=1$, and recall that $r_1\leq(n-2)$.
Possibly, $\ell_2={t}=n$.

Note that $\q(r_1)\geq0$ and $2\leq\q(\ell_2)\leq(k)$.

The $E_1$-edge $v_{(\ell_2-1)}v_{\ell_2}$ has capacity one or two.

If $r_1=s$,
then the $E_1$-edge $v_{r_1}v_{(r_1+1)}$ has capacity two,
otherwise, this edge has capacity one.

If $r_1\not=s$, then $\q(r_1)\geq1$;
moreover, $(r_1+1)\not\in{S}$ and $(r_1+1)$ is the first node of its $\Q$-set, $Q_{\q(r_1+1)}$;
hence, the $E_2$-edge between the last node of $\Q_{(\q(r_1)-1)}$ and $(r_1+1)$ is in $\delta(S)$.

At this point in the analysis, it is clear that $\delta(S)$ has capacity at least $3$.

We have two subcases, depending on $\q(\ell_2) - \q(r_1)$.

\begin{description}

\item[$\q(\ell_2) \geq \q(r_1)+3$:]
Then the interval $[(r_1+1),\dots,(\ell_2-1)]$ (which is a subset of $V-S$) contains two or more $\Q$-sets.
We will show that the second and the last-but-one of the $\Q$-sets each contribute
an $E_2$-edge to $\delta(S)$.

Since $r_1$ is the last node of its $\Q$-set, $r_1\in{S}$, and
the first node of $\Q_{(\q(r_1)+2)}$ is not in $S$,
the $E_2$-edge between these two nodes is in $\delta(S)$.

Similarly, since $\ell_2$ is the first node of its $\Q$-set, $\ell_2\in{S}$, and
the last node of $\Q_{(\q(\ell_2)-2)}$ is not in $S$,
the $E_2$-edge between these two nodes is in $\delta(S)$.

Thus, $\delta(S)$ has capacity $\geq5 =\lambda=5$.

\item[$\q(\ell_2) \leq \q(r_1)+2$:]

In this case, $\q(\ell_2) = \q(r_1)+2$, since the first~node of $\Q_{(\q(r_1)+1)}$,
which is $(r_1+1)$, is not in $S$, and $\ell_2$ (which is the first~node of its $\Q$-set) is in $S$.

Observe that $r_2\not=n$;
that is, $r_2\leq(n-1)$ and the node $(r_2+1)$ is not in $S$.
Otherwise, we would have $S=V-\Q_{(\q(r_1)+1)}$,
contradicting our assumption that $S$ is not the complement of a $\Q$-set.

First, suppose $r_2=(n-1)$.
Then $\delta(S)$ has the edge $v_{(n-1)}v_n$ of capacity two.
Hence, overall, $\delta(S)$ has capacity $\geq5 =\lambda=5$.

Now, suppose $r_2\leq(n-2)$.
If the $E_1$-edge $v_{r_2}v_{(r_2+1)}$ has capacity $3$,
then, overall, $\delta(S)$ has capacity $\geq6 =\lambda=5$.
Otherwise, the $E_1$-edge $v_{r_2}v_{(r_2+1)}$ has capacity one. Observe that
the node $r_2$ is the last~node of its $\Q$-set; moreover,
the node $(r_2+1)$ is the first~node of its $\Q$-set and $\q(r_2+1)\leq(k-1)$.
If the interval $[(r_2+1),(r_2+2),\dots,n]$ has one or more nodes of $S$,
then $\delta(S)$ has one more $E_1$-edge (besides $v_{r_2}v_{(r_2+1)}$) of capacity one or more,
hence, overall, $\delta(S)$ has capacity $\geq5 =\lambda=5$.
Finally, suppose that the interval $[(r_2+1),(r_2+2),\dots,n]$ has no nodes of $S$.
Then, the node $r_2$ is in $S$ and the first~node of $\Q_{(\q(r_2)+2)}$ is not in $S$,
hence, the $E_2$ edge between these two nodes is in $\delta(S)$.
Thus, overall, $\delta(S)$ has capacity $\geq5 =\lambda=5$.

\end{description}
}

\end{description}

\end{proof}
}

}
\end{appendices}

%%%%%%%%%%%%%%%%%%%%

\end{document}